\documentclass[prl,aps,showpacs,twocolumn,floatfix]{revtex4}
\usepackage{epsfig} \usepackage{graphics} \usepackage{bm}
\usepackage{amssymb}
\usepackage{graphicx}
\begin{document}

\preprint{Natala-PRB}

\title{Theory of Lee-Naughton-Lebed's
Oscillations in Moderately Strong Electric Fields in Layered
Quasi-One-Dimensional Conductors}

\author{A.G. Lebed$^*$}

\affiliation{Department of Physics, University of Arizona, 1118 E.
4-th Street, Tucson, AZ 85721, USA}

\begin{abstract}
In framework of some extension of the quasi-classical Boltzmann
kinetic equation, we show that a moderately strong electric field
splits the so-called Lee-Naughton-Lebed's magnetoconductivity
maxima in a layered quasi-one-dimensional conductor, if we use
some reasonable approximation to the equation. By means of the
above mentioned approximation, we obtain analytical formula for
conductivity in high magnetic and moderately high electric fields
and show that it coincides with the hypothetical formula as well
as adequately describes the pioneering experimental data by
Kobayashi et al. [K. Kobayashi, M. Saito, E. Omichi, and T. Osada,
Phys. Rev. Lett. \textbf{96}, 126601 (2006)].
\end{abstract}


\maketitle

In layered quasi-one-dimensional (Q1D) conductors in a magnetic
field, there are no closed orbits and, thus, the Landau
quantization is not possible. Nevertheless, there are other
quantum effects - the so-called Bragg reflections of electrons
from the Brillouin zones boundaries [1-4]. This leads to the
existence in (TMTSF)$_2$- and (ET)$_2$- based Q1D conductors of
such quantum phases as the Field-Induced-Spin(Charge)-Density-Wave
ones, exhibiting 3D Quantum Hall effect, and the Reentrant
Superconductivity (see, for the review, Ref.[4]). Metallic phases
of the above mentioned materials are also unusual and demonstrate
two original types of angular magnetic oscillations: the so-called
Lebed's magic angles (LMA) [5-28] and the Lee-Naughton-Lebed's
(LNL) oscillations [29-35]. As to the LMA effects, they still
contain lots of unexplained features and possibly have non
Fermi-Liquid origin [14,4,28], whereas the LNL oscillations are
well explained by present moment [33,36-42]. It is important that
the formulas for conductivity in regime of the LNL effects (see
Fig.1) are the same in quasi-classical extensions of the kinetic
equations [36,37] and in different pure quantum approaches
[33,34,38-42]. More recently Kobayashi et al. in the pioneering
work [43] have considered effects of moderately strong electric
fields on the LNL phenomenon and, in particular, have
experimentally shown that the strong electric field splits the LNL
maxima of conductivity. They have also theoretically suggested
some hypothetical formula for the LNL conductivity in a strong
electric field.

The goal of our paper is to show that the hypothetical formula of
Ref.[43] can be obtained by using some moderately high electric
field approximation for quasi-classical extension of the Boltzmann
kinetic equation. As shown below and as mentioned in Ref.[43], it
describes the splitting of the LNL conductivity maxima both at
qualitative and quantitative levels. As in Refs.[36,37], we use
periodic solutions of the Boltzmann kinetic equation in
$\tau$-approximation [44], which take into account quantum effects
of the Bragg reflections of electrons from the boundaries of the
Brillouin zones. Contrary to Refs.[36,37], we first keep in the
quasi-classical Boltzmann kinetic equation all terms,
corresponding to a strong electric field. Then, we use some
moderately strong electric field approximation and estimate where
we can neglect one of the above mentioned terms. As a result, we
obtain such moderately strong electric field approximation, which
differs from the results [36,37] and reproduce hypothetical
formula [43]. We discuss the applicability area of this formula
and show that it is broken in very high electric fields. We
demonstrate also that in moderately high electric fields it
describes splitting of the LNL maxima of conductivity as
experimentally observed in Ref. [43].
\begin{figure}[t]
\centering
\includegraphics[width=0.5\textwidth]{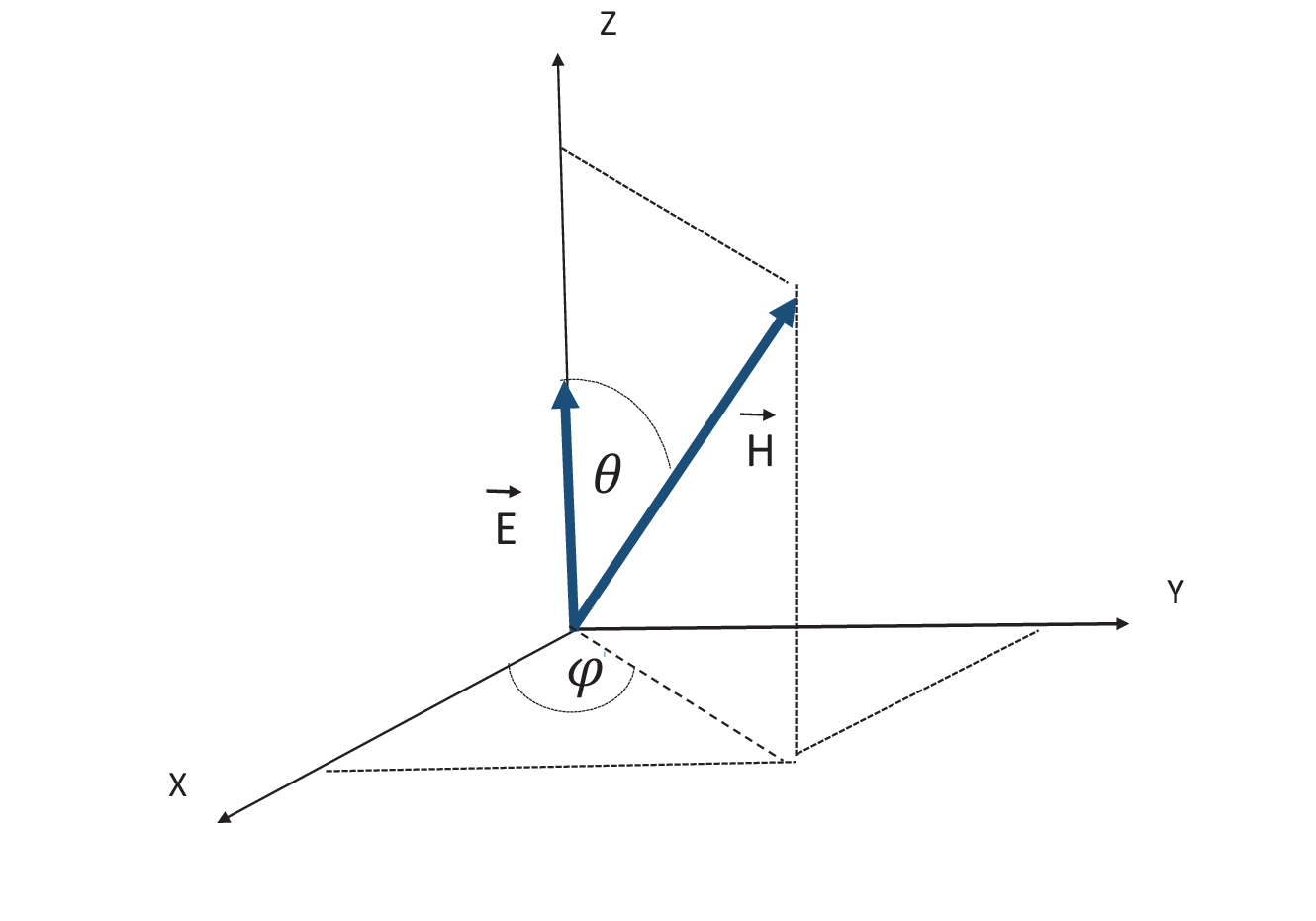}
\caption{In case of the LNL geometry, the direction of magnetic
field is characterized by two angles, $\theta$ and $\phi$, whereas
the electric field is applied perpendicular to the conducting
$(\bf{x},\bf{y})$ plane.}
\end{figure}

Let us consider the following Q1D Fermi surface in a layered
conductor in a tight-binding model:
\begin{eqnarray}
&\epsilon({\bf p})= \pm v_F (p_x \mp p_F) + 2 t_b \cos(p_y b^*) +
2 t_{\perp} \cos(p_z d_{\perp}),
\nonumber\\
&v_F p_F \gg t_b \gg t_{\perp},
\end{eqnarray}
where $v_F$ and $p_F$ are the Fermi velocity and Fermi momentum,
respectively; $t_b$ is the integral of overlapping of the electron
wave functions within the conducting plane, $t_{\perp}$ is the
integral of overlapping of the wave functions between the
conducting planes. Under the condition of the LNL experiment the
Q1D conductor is placed in the inclined magnetic field,
\begin{equation}
{\bf H} = H \ (\sin \theta \cos \phi, \sin \theta \sin \phi, \cos
\theta),
\end{equation}
whereas the constant electric field is applied perpendicular to
the conducting layers,
\begin{equation}
{\bf E} = E \ (0, 0, 1)
\end{equation}
(see Fig.1).

In the so-called $\tau$-approximation, the Boltzmann kinetic
equation can be written as [44]
\begin{equation}
\frac{d n({\bf p})}{dt}= - \frac{n({\bf p})-n_0({\bf p})}{\tau} \
,
\end{equation}
where $n({\bf p})$ is the electron distribution function and
$n_0({\bf p})$ is the Fermi-Dirac distribution function. In the
presence of external force, Eq.(4) can be rewritten as
\begin{equation}
{\bf F}({\bf p}) \frac{d n({\bf p})}{d {\bf p}} = - \frac{n({\bf
p})-n_0({\bf p})}{\tau},
\end{equation}
where the external force is
\begin{equation}
{\bf F}({\bf p}) =  e {\bf E}+\biggl( \frac{e}{c} \biggl) [{\bf
v(p)} \times {\bf H}],
\end{equation}
where $e$ and $c$ are the electron charge and the speed of light,
respectively; ${\bf v(p)}$ is the electron velocity. Thus, in the
presence of magnetic (2) and electric (3) forces, Eqs.(4)-(6) can
be represented as
\begin{equation}
\biggl\{ e{\bf E} + \biggl( \frac{e}{c} \biggl) [{\bf v(p)} \times
{\bf H}] \biggl\} \frac{d n({\bf p})}{d {\bf p}} = - \frac{n({\bf
p})-n_0({\bf p})}{\tau}.
\end{equation}
At low enough temperatures, $k_B T \ll \epsilon_F$, we introduce
as usual [44]:
\begin{equation}
n({\bf p})=n_0[\epsilon({\bf p})] - \frac{d n_0(\epsilon)}{d
\epsilon}\Psi ({\bf p}),
\end{equation}
where $\epsilon_F =v_F p_F$ is the Fermi energy, $k_B$ is the
Boltzmann constant.

As a result, we obtain for derivative of the electron distribution
function the following equation:
\begin{eqnarray}
\frac{d n({\bf p})}{d {\bf p}}=\frac{d n_0(\epsilon)}{d \epsilon}
\frac{d \epsilon ({\bf p})}{d {\bf p} }-\frac{d^2 n_0(\epsilon)}{d
\epsilon^2} \frac{d \epsilon ({\bf p})}{d {\bf p} }\Psi ({\bf p})
\nonumber\\
-\frac{d n_0(\epsilon)}{d \epsilon} \frac{d \Psi({\bf p})}{d \bf
p}.
\end{eqnarray}
If we take into account that in the quasi-classical approximation
\begin{equation}
\frac{d \epsilon({\bf p})}{d {\bf p}}={\bf v(p)},
\end{equation}
then we can rewrite Eq.(9) as
\begin{eqnarray}
\frac{d n({\bf p})}{d {\bf p}}= {\bf v({\bf p})} \biggl[\frac{d
n_0(\epsilon)}{d \epsilon} -\frac{d^2 n_0(\epsilon)}{d
\epsilon^2}\Psi ({\bf p}) \biggl]
\nonumber\\
-\frac{d n_0(\epsilon)}{d \epsilon} \frac{d \Psi({\bf p})}{d \bf
p}.
\end{eqnarray}
Using Eq.(11), we can now represent the quasi-classical Boltzmann
kinetic equation (7) in the following form:
\begin{eqnarray}
e{\bf E}{\bf v({\bf p})} \biggl[\frac{d n_0(\epsilon)}{d \epsilon}
-\frac{d^2 n_0(\epsilon)}{d \epsilon^2}\Psi ({\bf p}) \biggl]
\nonumber\\
-\biggl\{ e{\bf E} + \biggl( \frac{e}{c} \biggl) [{\bf v(p)}
\times {\bf H}] \biggl\}\frac{d n_0(\epsilon)}{d \epsilon} \frac{d
\Psi({\bf p})}{d \bf p}
 \nonumber\\
 = \frac{d n_0(\epsilon)}{d \epsilon} \frac{\Psi({\bf p})}{\tau}.
\end{eqnarray}

Note that the Boltzmann kinetic equation is usually studied in
metals in small electric fields, whereas the magnetic fields can
be strong. Therefore, there is usually considered a variant of the
equation, which is linear with respect to the electric field.
Since $\Psi({\bf p})$ and $d \Psi({\bf p})/d {\bf p}$ are both
proportional to electric field, the following two terms
\begin{equation}
-e{\bf E}{\bf v({\bf p})}\frac{d^2 n_0(\epsilon)}{d
\epsilon^2}\Psi ({\bf p}) -  e{\bf E} \frac{d n_0(\epsilon)}{d
\epsilon} \frac{d \Psi({\bf p})}{d \bf p}
\end{equation}
are usually omitted in the Boltzmann equation (12) (see, for
example, Refs.[36,37]). In this article, for the first time we
theoretically consider the case of moderately strong electric
fields, where we disregard the first term but keep the second one
of the above mentioned two terms (13). It is easy to see that we
can disregard the first term in Eq.(13), if it it much less than
the right side of Eq.(12):
\begin{equation}
\biggl| e{\bf E}{\bf v({\bf p})}\frac{d^2 n_0(\epsilon)}{d
\epsilon^2} \Psi({\bf p}) \biggl| \ll \biggl| \frac{d
n_0(\epsilon)}{d \epsilon} \frac{\Psi({\bf p})}{\tau} \bigg|.
\end{equation}
Since
\begin{equation}
|{\bf v}({\bf p})| = |-2t_{\perp} d_{\perp} \sin(p_z d_{\perp})|
\sim t_{\perp} d_{\perp}
\end{equation}
and
\begin{equation}
 \biggl| \frac{d^2 n_0(\epsilon)}{d
\epsilon^2} \biggl| \sim \frac{1}{T} \biggl| \frac{d
n_0(\epsilon)}{d \epsilon} \biggl|,
\end{equation}
Eq.(14) can be rewritten as
\begin{equation}
 eE(t_{\perp} d_{\perp}) \tau \ll T.
\end{equation}
The physical meaning of Eqs.(14)-(17) is now clear. Electric field
has to be small enough in order not to change electron energy on
the scale of the temperature. As a result of disregarding the
above discussed term in Eq.(12), instead of Eq.(12), we obtain
\begin{equation}
e{\bf E}{\bf v({\bf p})} -\biggl\{ e{\bf E} + \biggl( \frac{e}{c}
\biggl) [{\bf v(p)} \times {\bf H}] \biggl\} \frac{d \Psi({\bf
p})}{d \bf p}
 =  \frac{\Psi({\bf p})}{\tau}.
\end{equation}
It is important that Eq.(18) is different from the weak electric
field approximation equations considered in Refs. [36,37] and,
thus, we call the former equation the quasi-classical kinetic
equation for moderately strong electric fields.

Let us now take into account the layered Q1D nature of the
electron spectrum (1) placed in the magnetic field (2) and the
electric field (3). In this case, we can disregard the Lorentz
force component in Eq.(18), originated from velocity component
along ${\bf z}$ axis ($t_c \ll t_b$), and obtain the following
kinetic equation near right sheet of the Q1D FS (where $v_x
\approx + v_F$):
\begin{eqnarray}
-e E v^0_z \sin (\tilde z) + \omega_b (\theta) \frac{\partial
\Psi^+(\tilde y, \tilde z)}{\partial \tilde y} -
\omega^+_c(\theta,\phi) \frac{\partial \Psi^+(\tilde y, \tilde
z)}{\partial \tilde z}
\nonumber\\
- \omega^*_c(\theta,\phi) \sin(\tilde y) \frac{\partial
\Psi^+(\tilde y, \tilde z)}{\partial \tilde z}
 =  \frac{\Psi^+(\tilde y, \tilde z)}{\tau}.
\end{eqnarray}
Note that near left sheet of the layered Q1D FS (1) we obtain a
slightly different equation:
\begin{eqnarray}
-e E v^0_z \sin (\tilde z) - \omega_b (\theta) \frac{\partial
\Psi^-(\tilde y, \tilde z)}{\partial \tilde y} +
\omega^-_c(\theta,\phi) \frac{\partial \Psi^-(\tilde y, \tilde
z)}{\partial \tilde z}
\nonumber\\
- \omega^*_c(\theta,\phi) \sin(\tilde y) \frac{\partial
\Psi^-(\tilde y, \tilde z)}{\partial \tilde z}
 =  \frac{\Psi^-(\tilde y, \tilde z)}{\tau}.
\end{eqnarray}
Let us specify notations used in Eqs.(19) and (20):
\begin{equation}
v^0_z = 2 t_{\perp} d_{\perp} , \ v^0_y = 2 t_b b^*, \ \ \  \tilde
y = p_y b^*, \ \tilde z = p_z d_{\perp}
\end{equation}
and
\begin{eqnarray}
\omega_b (\theta)= e v_F b^* \cos (\theta)H/c, \ \
\omega^{\pm}_c(\theta,\phi)=\omega_c(\theta,\phi) \pm \omega_E,
\nonumber\\
\omega_c(\theta,\phi)=e v_F d_{\perp} \sin(\theta) \sin(\phi) H/c
, \ \ \omega_E=eEd_{\perp},
\nonumber\\
\omega^*_c(\theta,\phi) = e v^0_y d_{\perp} \sin(\theta)
\cos(\phi) H/c.
\end{eqnarray}

It is important that Eqs.(19) and (20) can be solved analytically.
As a result of lengthly but rather straightforward calculations,
we obtain
\begin{eqnarray}
\Psi^+(\tilde y, \tilde z) = -\frac{e E v^0_z}{\omega_b (\theta)}
\int^{\infty}_{\tilde y} \sin \biggl\{ (\tilde z) +
\frac{\omega^+_c(\theta,\phi)}{\omega_b (\theta)}(\tilde y -t)
\nonumber\\
+ \frac{\omega^*_c(\theta,\phi)}{\omega_b (\theta)} \biggl[
\cos(t)-\cos(\tilde y) \biggl] \biggl\} \exp \biggl[-\frac{t-
\tilde y}{\tau \omega_b (\theta)} \biggl] \ dt
\end{eqnarray}
and
\begin{eqnarray}
\Psi^-(\tilde y, \tilde z) = -\frac{e E v^0_z}{\omega_b (\theta)}
\int^{\tilde y}_{-\infty} \sin \biggl\{(\tilde z) +
\frac{\omega^-_c(\theta,\phi)}{\omega_b (\theta)}(\tilde y -t)
\nonumber\\
- \frac{\omega^*_c(\theta,\phi)}{\omega_b (\theta)} \biggl[
\cos(t)-\cos(\tilde y) \biggl] \biggl\} \exp \biggl[\frac{t-
\tilde y}{\tau \omega_b (\theta)} \biggl] \ dt.
\end{eqnarray}

It is easy to understand that the total current can be written as
a summation of two currents: one from the right and another from
the left sheets of the Q1D FS (1),
\begin{equation}
j_z (E,{\bf H}) = j^+_z (E,{\bf H}) + j^-_z (E,{\bf H}),
\end{equation}
which are proportional to:
\begin{eqnarray}
j^+_z(E,{\bf H}) \sim e(v^0_z)\biggl[\frac{e E v^0_z}{\omega_b
(\theta)}\biggl] \int^{\pi}_{-\pi} \frac{d \tilde z}{2 \pi}
\int^{\pi}_{-\pi} \frac{d \tilde y}{2 \pi} \sin(\tilde z)
\nonumber\\
\times \int^{\infty}_{\tilde y} \sin \biggl\{(\tilde z) +
\frac{\omega^+_c(\theta,\phi)}{\omega_b (\theta)}(\tilde y -t)
\nonumber\\
+ \frac{\omega^*_c(\theta,\phi)}{\omega_b (\theta)} \biggl[
\cos(t)-\cos(\tilde y) \biggl] \biggl\} \exp \biggl[-\frac{t-
\tilde y}{\tau \omega_b (\theta)} \biggl] \ dt
\end{eqnarray}
and
\begin{eqnarray}
j^-_z(E,{\bf H}) \sim  e(v^0_z)\biggl[\frac{e E v^0_z}{\omega_b
(\theta)}\biggl] \int^{\pi}_{-\pi} \frac{d \tilde z}{2 \pi}
\int^{\pi}_{-\pi} \frac{d \tilde y}{2 \pi} \sin(\tilde z)
\nonumber\\
\times \int^{\tilde y}_{-\infty} \sin \biggl\{(\tilde z) +
\frac{\omega^-_c(\theta,\phi)}{\omega_b (\theta)}(\tilde y -t)
\nonumber\\
- \frac{\omega^*_c(\theta,\phi)}{\omega_b (\theta)} \biggl[
\cos(t)-\cos(\tilde y) \biggl] \biggl\} \exp \biggl[\frac{t-
\tilde y}{\tau \omega_b (\theta)} \biggl] \ dt .
\end{eqnarray}
From Eqs.(26) and (27), it follows that
\begin{eqnarray}
\sigma^+_{zz}(E,{\bf H}) = \frac{\sigma_{zz}(0)}{\omega_b
(\theta)\tau } \int^{\pi}_{-\pi} \frac{d \tilde y}{2 \pi}
\int^{\infty}_{\tilde y} \cos \biggl\{
\frac{\omega^+_c(\theta,\phi)}{\omega_b (\theta)}(\tilde y -t)
\nonumber\\
+ \frac{\omega^*_c(\theta,\phi)}{\omega_b (\theta)} \biggl[
\cos(t)-\cos(\tilde y) \biggl] \biggl\} \exp \biggl[-\frac{t-
\tilde y}{\tau \omega_b (\theta)} \biggl] \ dt
\end{eqnarray}
and
\begin{eqnarray}
\sigma^-_{zz}(E,{\bf H}) = \frac{\sigma_{zz}(0)}{\omega_b
(\theta)\tau} \int^{\pi}_{-\pi} \frac{d \tilde y}{2 \pi}
\int^{\tilde y}_{-\infty} \cos \biggl\{
\frac{\omega^-_c(\theta,\phi)}{\omega_b (\theta)}(\tilde y -t)
\nonumber\\
- \frac{\omega^*_c(\theta,\phi)}{\omega_b (\theta)} \biggl[
\cos(t)-\cos(\tilde y) \biggl] \biggl\} \exp \biggl[\frac{t-
\tilde y}{\tau \omega_b (\theta)} \biggl] \ dt ,
\end{eqnarray}
where
\begin{equation}
\sigma_{zz}(E,{\bf H})= \sigma^+_{zz}(E,{\bf H}) +
\sigma^-_{zz}(E,{\bf H}).
\end{equation}
Note that in Eqs.(28)-(30), the electric and magnetic field
dependent conductivity is defined as
\begin{equation}
j_{zz}(E,{\bf H}) = E \sigma_{zz}(E,{\bf H}), \ \
\sigma_{zz}(0)=\sigma_{zz}(E=0,{\bf H}=0).
\end{equation}
Straightforward calculations of the integrals in Eqs. (28) and
(29) result in the following expression for the total conductivity
(30):
\begin{eqnarray}
\sigma_{zz}(\theta, \phi,E,H) = \sigma_{zz}(E,{\bf H}) =
\frac{\sigma_{zz}(0)}{2} \sum^{+ \infty}_{n=-\infty} J^2_n
\biggl[\frac{\omega^*_c(\theta,\phi)}{\omega_b (\theta)} \biggl]
\nonumber\\
\times \biggl\{\frac{1}{1+ [\omega_c(\theta,\phi)+ \omega_E - n
\omega_b (\theta)]^2 \tau^2 }
\nonumber\\
+ \frac{1}{1+ [\omega_c(\theta,\phi)- \omega_E - n \omega_b
(\theta)]^2 \tau^2} \biggl\}.
\end{eqnarray}
Note that Eq.(32) describes splitting of the LNL maxima of
conductivity for the LNL oscillations (see Fig.2 of Ref.[43]).
Indeed, in a pure layered Q1D metals it has two maxima at
\begin{equation}
\omega_c(\theta,\phi)= n \omega_b(\theta) \pm  \omega_E
\end{equation}
or
\begin{equation}
\tan(\theta^{\pm}) \sin(\phi)= n \biggl( \frac{b^*}{d_{\perp}}
\biggl) \pm \frac{E c}{v_F H \cos(\theta)},
\end{equation}
where $n$ is an integer. We note that, using Eq.(34) and
experimental data on splitting the LNL maxima, the authors of work
[43] evaluated the Fermi velocity $v_F$ (1) in compound
$\alpha$-(BEDT-TTF)$_2$KHg(SCN)$_4$, corresponding to open sheets
of the Fermi surface, $v_F \simeq 10^7$cm/s. We suggest to use the
above described effect to determine Fermi velocities in other Q1D
conductors, where heating of a sample under experiment allows to
observe such splitting and where inequality (17) is fulfilled.

To summarize we stress that the derived above in moderately high
electric fields (i.e., when inequality (17) is fulfilled) Eq.(32)
was guessed in Ref.[43] as a strict equation, which is not
correct. Although Eq.(32) coincides with Eq.(4) from Ref. [43], we
have to check if inequality (17) is true under the experimental
conditions of Ref.[43]. Indeed, the experimental conditions were
the following: voltage $V$ = 2-20 V, thickness of the sample $d
=0.1$ mm, temperature $T=1.8$ K [43]. If we take into account the
following band structure parameters of
$\alpha$-(BEDT-TTF)$_2$KHg(SCN)$_4$ organic material [43]:
$d_{\perp}=20 \AA$ [4] and $t_{\perp} \simeq 30$ $\mu$eV [45],
then at $V$=2 V, Eq.(17) can be written as
\begin{equation}
 eE(t_{\perp} d_{\perp}) \tau \simeq 0.14 \ K \ll \ T = 1.8 \ K,
\end{equation}
whereas at $V$ = 20 V both sides of Eq.(17) become of the same
order. So, although the overall comparison of the experimental
results [43] with the theoretical Eq.(32) can be justified at
small voltages, at high voltages this has to be done with some
caution. In conclusion, we demonstrate equation showing how
Eq.(17) limits area for application of the Eq.(34) to describe the
LNL maxima splitting:
\begin{equation}
\tan(\theta^+)-\tan(\theta^-) \ll \frac{2T}{t_{\perp}}
\frac{\tan(\theta)}{\omega_c(\theta,\phi)\tau} .
\end{equation}

The author is thankful to N.N. Bagmet(Lebed) for useful
discussions.

$^*$Also at: L.D. Landau Institute for Theoretical Physics, RAS, 2
Kosygina Street, Moscow 117334, Russia.


\begin{references}

\bibitem{Lebed-1}
L.P. Gor'kov and A.G. Lebed, J. Phys. (Paris) Lett. \textbf{45},
L-433 (1984).

\bibitem{Heritier-1}
M. Heritier, G. Montambaux, and P. Lederer, J. Phys. (Paris) Lett.
\textbf{45}, L-943 (1984).

\bibitem{Chaikin-1}
P.M. Chaikin, Phys. Rev. B \textbf{31}, 4770 (1985).

\bibitem{Lebed-2}
A.G. Lebed ed., {\it The Physics of Organic Superconductors and
Conductors} (Springer-Verlag, Berlin, 2008).

\bibitem{Naughton-1}
M.J. Naughton, O.H. Chung, L.Y. Chiang, and J.S. Brooks, Material
Research Society Symposium Proceedings, \textbf{173}, 257 (1990).

\bibitem{Osada-1}
T. Osada, A. Kawasumi, S. Kagoshima, N. Miura, and G. Saito, Phys.
Rev. Lett. \textbf{66}, 1525 (1991).

\bibitem{Boebinger} G. S. Boebinger, G. Montambaux, M. L. Kaplan, R. C. Haddon, S.
V. Chichester, L. Y. Chiang, Phys. Rev. Lett. \textbf{64}, 591
(1990).

\bibitem{Naughton-2}
M. J. Naughton, O. H. Chung, M. Chaparala, X. Bu, P. Coppens,
Phys. Rev. Lett. \textbf{67}, 3712 (1991).

\bibitem{Chaikin-2}
W. Kang, S. T. Hannahs, and P. M. Chaikin, Phys. Rev. Lett.
\textbf{69}, 2827 (1992).

\bibitem{Karts-1}
M.V. Kartsovnik, A.E. Kovalev, V.N. Laukhin, and S.I. Pesotskii,
J. Phys. I (France) \textbf{2}, 223 (1992).

\bibitem{Karts-2}
M.V. Kartsovnik, A.E. Kovalev, and N.D. Kushch, J. Phys. I
(France) \textbf{3}, 1187 (1993).

\bibitem{Benhia}
K. Benhia, M. Ribault, and C. Lenior, Europhys. Lett. \textbf{25},
285 (1994).

\bibitem{Karts-2}
M.V. Kartsovnik, A.E. Kovalev, V.N. Laukhin, H. Ito, T. Ishiguro,
N.D. Kushch, H. Anzai, and G. Saito, Synth. Met. \textbf{70}, 819
(1995).

\bibitem{Chaikin-3}
E.I. Chashechkina and P.M. Chaikin, Phys. Rev. Lett. \textbf{80},
2181 (1998).

\bibitem{Osada-2}
T. Osada, H. Nose, and Kuraguchi, Physica B \textbf{294-295}, 402
(2001).

\bibitem{Chaikin-4}
E.I. Chashechkina and P.M. Chaikin, Phys. Rev. B \textbf{65},
012405 (2002).

\bibitem{Kang-1}
H. Kang, Y.J. Jo, S. Uji, and W. Kang, Phys. Rev. B \textbf{68},
132508 (2003).

\bibitem{Kang-2}
H. Kang, Y.J. Jo, and W. Kang, Phys. Rev. B \textbf{69}, 033103
(2004).

\bibitem{Ito-1}
H. Ito, D. Suzuki, Y. Yokochi, S. Kuroda, M. Umemiya, H. Miyasaka,
K-I. Sugiura, M. Yamashita, H. Tajima, Phys. Rev. B \textbf{71},
212503 (2005).

\bibitem{Karts-3}
M. V. Kartsovnik, D. Andres, S. V. Simonov, W. Biber- acher, I.
Sheikin, N. D. Kushch, and H. Miller, Phys. Rev. Lett.
\textbf{96}, 166601 (2006).

\bibitem{Hill-1}
S. Takahashi, A. Betancur-Rodiguez, S. Hill, S. Takasaki, J.
Yamada and H. Anzai, J. Low Temp. Phys. \textbf{142}, 315 (2007).

\bibitem{Kang-3}
W. Kang, T. Osada, Y.J. Jo, and Haeyong Kang, Phys. Rev. Lett.
\textbf{99}, 017002 (2007).

\bibitem{Kang}
W. Kang, Phys. Rev. B \textbf{76}, 193103 (2007).

\bibitem{Singleton-1}
A.F. Bangura, P.A. Goddard, J. Singleton, S.W. Tozer, A.I. Coldea,
A. Ardavan, R.D. McDonald, S.J. Blundell and J.A. Schlueter, Phys.
Rev. B \textbf{76}, 0525010 (2007).

\bibitem{Kang-5}
W. Kang, Ok-Hee Chung, Phys. Rev. B \textbf{79}, 045115 (2009).

\bibitem{Brooks-1}
D. Graf, J.S. Brooks, E.S. Choi, M. Almeida, R.T. Hen- riques,
J.C. Dias, and S. Uji, Phys. Rev. B \textbf{80}, 155104 (2009).

\bibitem{Kang-6}
W. Kang, Y.J. Jo, D.Y. Noh, K.I. Son, and Ok-Hee Chung, Phys. Rev.
B \textbf{80}, 155102 (2009).

\bibitem{Uji}
Kaya Kobayashi, H. Satsukawa, J. Yamada, T. Terashima, and S. Uji,
Phys. Rev. Lett. \textbf{112}, 116805 (2014).


\bibitem{Naughton-4}
M.J. Naughton, I.J. Lee, P.M. Chaikin, and G.M. Danner, Synth.
Metals \textbf{85}, 1481 (1997).

\bibitem{Yoshino}
H. Yoshino, K. Saito, H. Nishikawa, K. Kikuchi, K. Kobayashi, and
I. Ikemoto, J. Phys. Soc. Jpn. \textbf{66}, 2248 (1997).

\bibitem{Lee-1}
I.J. Lee and M.J. Naughton, Phys. Rev. B \textbf{57}, 7423 (1998).

\bibitem{Lee-2}
I.J. Lee and M.J. Naughton, Phys. Rev. B \textbf{58}, R13343
(1998).

\bibitem{Lebed-6}
A.G. Lebed and M.J. Naughton, Phys. Rev. Lett. \textbf {91},
187003 (2003).

\bibitem{Lebed-7}
A.G. Lebed, Heon-Ick Ha, and M.J. Naughton, Phys. Rev. B
\textbf{71}, 132504 (2005).

\bibitem{Lebed-8}
H.I. Ha, A.G. Lebed, and M.J. Naughton, Phys. Rev. B \textbf{73},
033107 (2006).

\bibitem{McKenzie}
R.H. McKenzie and P. Moses, Phys. Rev. B \textbf{60}, R11241
(1999).

\bibitem{Lebed-9}
A.G. Lebed and M.J. Naughton, J. Phys. IV (France) \textbf{12},
369 (2002).

\bibitem{Osada-2}
T. Osada, Physica E \textbf{12}, 272 (2002).

\bibitem{Osada-3}
T. Osada and M. Kuraguchi, Synth. Met. \textbf{133-134}, 75
(2003).

\bibitem{Yakovenko-1}
A. Banerjee and V. Yakovenko, Phys. Rev. B \textbf{78}, 125404
(2008).

\bibitem{Yakovenko-2}
B. K. Cooper, V. M. Yakovenko, Phys. Rev. Lett. \textbf{96},
037001 (2006).


\bibitem{Lebed-9}
S. Wu and A.G. Lebed, Phys. Rev. B \textbf{82}, 075123 (2010).

\bibitem{Kobayashi}
K. Kobayashi, M. Saito, E. Omichi, and T. Osada, Phys. Rev. Lett.
\textbf{96}, 126601 (2006).


\bibitem{Abrikosov-1}
See, for example, book A.A. Abrikosov, {\it Fundamentals of Theory
of Metals} (Elsevier Science, Amsterdam, 1988).

\bibitem{Grigoriev}
P.D. Grigoriev, M.V. Kartsovnik, and W. Biberacher, Phys. Rev. B
\textbf{86}, 165125 (2012).










\end{references}
\end{document}